# Characterization of a Diode Dosimeter for UHDR FLASH Radiotherapy

Mahbubur Rahman[1,2*], Jakub Kozelka[3], Jeff Hildreth[3], Andreas Schönfeld[3], Austin M. Sloop[1], M. Ramish Ashraf[1,4], Petr Bruza[1], David J. Gladstone[1,5,6], Brian W. Pogue[1,6,7,8], William E. Simon[2], Rongxiao Zhang[1,4,5*]

[1] Thayer School of Engineering, Dartmouth College, Hanover NH 03755, US
[2] UT Southwestern Medical Center, Dallas TX 75235, US
[3] Sun Nuclear Inc., Melbourne FL 32940, US
[4] Stanford University, Stanford, CA 94305, US
[5] Department of Medicine, Radiation Oncology, Geisel School of Medicine, Dartmouth College Hanover NH 03755 USA
[6] Dartmouth Cancer Center, Dartmouth-Hitchcock Medical Center, Lebanon, NH 03756 USA
[7] Department of Surgery, Geisel School of Medicine, Dartmouth College, Hanover NH 03755 USA
[8] Department of Medical Physics, Wisconsin Institutes for Medical Research, University of Wisconsin, Madison WI 53705 USA

[*] Correspondence to: Mahbubur.Rahman@UTSouthwestern.edu or Rongxiao.Zhang@hitchcock.org

**Abstract:**

**Background:** Ultra-high dose rate (UHDR) FLASH beams typically deliver dose at rates of > 40Gy/sec. Characterization of these beams with respect to dose, mean dose rate, and dose per pulse requires dosimeters which exhibit high temporal resolution and fast readout capabilities.

**Purpose:** A diode EDGE Detector with a newly designed electrometer has been characterized for use in an UHDR electron beam and demonstrated appropriateness for UHDR FLASH radiotherapy dosimetry.

**Methods:** Dose linearity, mean dose rate, and dose per pulse dependencies of the EDGE Detector were quantified and compared with dosimeters including a W1 scintillator detector, radiochromic film, and ionization chamber that were irradiated with a 10 MeV UHDR beam. The dose, dose rate and dose per pulse were controlled via an in-house developed scintillation-based feedback mechanism, repetition rate of the linear accelerator, and source-to-surface distance, respectively. Depth-dose profiles and temporal profiles at individual pulse resolution were compared to the film and scintillation measurements, respectively. The radiation-induced change in response sensitivity was quantified via irradiation of ~5kGy.

**Results:** The EDGE Detector agreed with film measurements in the measured range with varying dose (up to 70 Gy), dose rate (nearly 200 Gy/s), and dose per pulse (up to 0.63 Gy/pulse) on average to within 2%, 5%, and 1%, respectively. The detector also agreed with W1 scintillation detector on average to within 2% for dose per pulse (up to 0.78 Gy/pulse). The EDGE Detector signal was proportional to ion chamber (IC) measured dose, and mean dose rate in the bremsstrahlung tail to within 0.4% and 0.2% respectively. The EDGE Detector measured percent depth dose agreed with film to within 3% and per pulse output agreed with W1 scintillator to within – 6% to +5%. The radiation-induced response decrease was 0.4% per kGy.

**Conclusions:** The EDGE Detector demonstrated dose linearity, mean dose rate independence, and dose per pulse independence for UHDR electron beams. It can quantify the beam spatially, and temporally at sub millisecond resolution. It's robustness and individual pulse detectability of treatment deliveries can potentially lead to its implementation for *in vivo* FLASH dosimetry, and dose monitoring.

## 1. Introduction

Recent advances in FLASH radiotherapy from delivery of ultra-high dose rate (UHDR, >40Gy/s) beams has further expanded the requirements of dosimetry tools to provide accurate dose delivered during treatments. The FLASH effect, or reduction in normal tissue damage with equivalent tumor control, was demonstrated *in vivo* in both small animals[1–5], large animals[6], and human treatment[7]. UHDR beam deliveries in preclinical and clinical studies typically occur in less than a second, requiring dose, mean dose rate, and dose per pulse independent dosimeters in the FLASH regime (>40 Gy/s, >0.1 Gy/pulse, >$10^5$ Gy/s intrapulse) with high temporal resolution[8] and fast readout. Typical clinical accelerators modified for UHDR delivery, like the one utilized in this study, have a pulse width and repetition rate (period) of 4 microseconds and 360Hz (2.78 milliseconds) respectively.

Currently there are several types of dosimeters utilized for characterizing UHDR beams and its dosimetry. Ion chambers (IC), a charge-based detector considered the gold standard in dosimetry due to their consistency and reliability, can provide real time monitoring of the beamline albeit requiring correction for ion recombination effects and reduced response at UHDR[9,10]. Alanine, and film[11,12] chemical-based detectors, and thermo-luminescent dosimeters[13] (TLD's) have been utilized for *in vivo* dosimetry but lack temporal resolution and require processing and read out post irradiation. Gafchromic film has also been shown to measure beam profiles due to their submillimeter spatial resolution[14]. Optically stimulated luminescence dosimeters (OSLDs), typically a dose rate independent, passive and post irradiation read out dosimeter[8], have progressed to provide real time beam characterization, but require further validation under UHDR conditions[15,16]. Alternatively, scintillation and Cherenkov emission-based detectors can provide real time dose measurements at millimeter and microsecond resolution[17,18]. Nonetheless there is a lack of studies investigating the use of solid-state detectors for FLASH dosimetry. Although, recently diamond detectors demonstrated dose rate independence for synchrotron x-ray beams up to 700 Gy/s[19] and electron beams at dose per pulse up to 6.5Gy[20]. Diode detectors have often been used to characterize conventional external beam irradiators due to their real time read out, high sensitivity, reliability and robustness[21]. Diodes are a viable option for small field dosimetry including IMRT plans with small beamlets, due to their small radiation sensitive volume[22].

In this study, an EDGE Detector (Sun Nuclear Corp, Melbourne, FL) with a modified electrometer was characterized in UHDR electron beams with regards to dose linearity, mean dose rate dependence, and dose per pulse dependence of the detector response. A W1 scintillator detector[23] studied[23,24] at conventional and UHDR conditions respectively (Standard Imaging, Middleton, WI), and EBT-XD Gafchromic film studied[25] at UHDR conditions (Ashland Advanced Materials, Bridgewater NJ), have been demonstrated to be dose rate independent in previous studies, and an IC (Exradin A28, Standard Imaging, Middleton, WI) served as reference. The IC was positioned in the bremsstrahlung tail of the beam to circumvent its dose rate dependence. The beam's depth and temporal profiles were then characterized with the EDGE Detector and compared with film and W1 detectors, respectively. The potential utility of the EDGE Detector under UHDR FLASH conditions is discussed.

## 2. Materials and Methods

### 2.A Description of EDGE Detector and Electrometer

#### 2.A. i. EDGE Detector

The commercially available SNC model 1118 EDGE Detector is an N type diffused junction silicon diode with a nominal sensitivity of 30 nC/Gy. It is encased in a brass housing (external dimensions of 3.8×5.5×38 mm$^3$) with an entrance window centered over the active junction of 0.8 x 0.8 mm². Detector assembly is terminated with a triaxial connector, with the center pin to the junction, middle guard shield to the die base, outer braid to the brass housing.

**2.A. ii. Electrometer**

The PC electrometer (PCE, SNC Model 1014) functions with the operational amplifier (OP-AMP) input connected to the radiation detector, with the OP-AMP feedback capacitor connected between the output and input of the OP-AMP. A 16-bit bi-polar Analog to Digital Converter (A2D) samples the output of the OP-AMP every 100 microseconds, with data decimation ability of M, that is, keep only every Mth sample, where M can take on a value of 1, 2, 4, 8, etc. Its voltage resolution is 0.15259 mV per A2D count and charge resolution is 15.3 pC per A2D count. Between each A2D measurement, a microcontroller determines LINAC pulse occurrence by A2D interval subtraction and initiates a charge pump when required, based upon voltage level near bottom or top of amplifier range. For this application, the modifications to the PCE by the manufacturer included an increase of amplifier feedback capacitor to accommodate increase in charge from EDGE, decrease the series input resistance and add an input buffer capacitor to provide a time constant for total charge transfer between LINAC pulses to the amplifier, decrease the resistor in the bucket pump to discharge the larger feedback capacitor, and an adjustment of temporal controls for the charge pump in the PCE. The input buffer circuit of 2 kΩ resistance and 100 nF capacitance combine to give a time constant, RxC, of 200 μs and a 10% to 90% rise time, 2.2xRxC, of 440 μs.

**2.B. Setup**

A clinical LINAC (Varian Clinac 2100 C/D) was converted to deliver UHDR 10 MeV electron beams at the isocenter by retracting the x-ray target and flattening filter and letting the electron beam normally intended to produce photons pass through the beamline[26]. Prior to conducting experiments, consistency of the beam dose per pulse and spatial distribution (shift, FWHM, flatness, symmetry) was verified via optical imaging of a scintillation screen on a solid water phantom[18]. A W1 point scintillator detector was coupled to a gated integrating amplifier and a real-time controller for dose monitoring and feedback control loop[24]. The controller was programmed to integrate dose per pulse when the sync pulse measurement exceeded a threshold and provide feedback to the LINAC when the prescribed dose was delivered, which stopped the beam after delivery of desired dose through a reed relay connected to a real time position management gating switchbox (Varian Inc, CA).

For comparison of the EDGE Detector to W1 and radiochromic film, the experimental setup shown in **Figure 1a** was used. For each radiation delivery, the response of the W1 scintillator and the EDGE Detector were measured with the cumulative dose recorded by EBT-XD Gafchromic film. The W1 scintillation detector was in between a 1 cm thick solid water phantom (superficial) and 1.15 cm thick solid water phantom (Gammex Model 457, Sun Nuclear) with the end of the fiber aligned to the isocenter. The EDGE detector was below the 2.15 cm of solid water phantom with another 5 cm of the solid water (Solid Water HE, Sun Nuclear) to provide full backscatter (for both the EDGE and W1). Note for percent depth dose (PDD) measurements solid water phantom thickness above the EDGE detector varied, which is described further at the end of this section. Both the EDGE and W1 detectors were in their respective machined slots in the solid water phantom. The film, flattened on top of the phantom, measured dose at the surface. The uncertainty in dose at the surface, determined from a single delivery to several adjacent film at the surface, was measured to be within 1.2% of the mean delivered dose. The calibration of the W1 scintillation was based on the manufacturer's suggestions for determining calibration coefficients (gain and Cherenkov to light ratio (CLR))[24]. This setup was used to measure the detectors' response with respect to varying mean dose rate, dose, dose per pulse, and per pulse beam output. The mean dose rate was varied by changing the repetition rate of the LINAC, while the dose per pulse was varied by changing the source to surface distance. The dose per pulse variation also occurred due to the ramp-up pulses in the LINAC, which allowed for comparison of individual pulses responses between W1 and EDGE detectors. The dose was varied by utilizing the beam stopping mechanism integrated with the W1 scintillator[24].

For comparison of the EDGE Detector to IC the setup shown in **Figure 1b** was used, again the EDGE and IC measured output simultaneously during each delivery from the LINAC. The IC was in the bremsstrahlung tail (17 cm depth) of the electron beam to mitigate potential ion recombination effects from UHDR beams[9]. According to the Varian Eclipse beam model of the UHDR 10 MeV electron beam[27], at 17 cm depth, the dose and dose rate is ~0.1% that of $d_{max}$ ensuring conventional dose rates for the IC. The IC correction factors for charge collection efficiency ($P_{ion}$) and polarity ($P_{pol}$) were measured/verified to be near unity with bias voltages at 300V/150V and 300V/-300V respectively. The IC response at 17 cm depth was calibrated to the dose measured with radiochromic film at 2 cm depth (Figure 1b). The calibration was then utilized to display the comparison of the EDGE Detector's response in UHDR conditions and the IC response in the bremsstrahlung tail of the electron beam in terms of dose for all future measurements, which included characterizing the EDGE Detector's response to dose, mean dose rate, and long-term stability. Note that for all dose and dose rate measurements it was only necessary to confirm a constant ratio between the two compared devices (including W1 and film). Furthermore, the output per pulse comparison between W1[23] and EDGE detector was measured simultaneously. The long-term stability of the EDGE detector sensitivity to dose was evaluated based on 500 deliveries, each with an average dose of 10.5 Gy. Its stability is compared to the W1[23], included in the discussion.

The PDD curve required additional solid water build-up of varying thickness, which were placed in between the 1.15 cm slab and the EDGE Detector. So, the depths of the EDGE detector ranged from 1.15 cm to 6.55 cm. The source to surface distance of 100 cm and the alignment of the detectors with the central axis of the beam were maintained in the process. The film measurements for comparison were completed separately but same number of pulses (35) were delivered, and results were normalized to the same depth of maximum dose (2.6 cm).

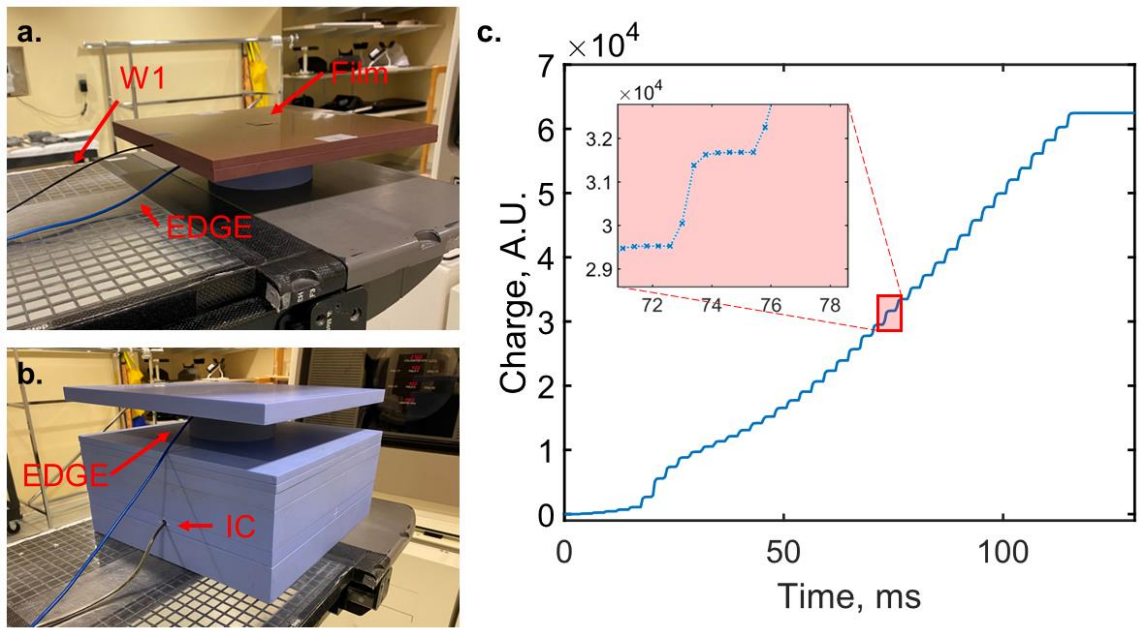

**Figure 1. a.** Setup for data acquisition from both the W1 detector and characterization of the EDGE Detector. Film on top of the stack of solid water phantom was used for calibration of the W1 detector and verifying dose for each delivery. **b.** Set up for data acquisition from both IC in the Bremsstrahlung tail (17 cm depth) and EDGE Detector (both calibrated to dose measured at 2cm depth from film. **c.** Example Snapshot data acquired from the EDGE Detector with a zoomed in plot of charge accumulation from an example individual pulse at 0.4 ms sampling rate. Data acquisition is not synchronous to pulse timing. In the zoomed in plot, the pulse occurred approximately 0.06 ms prior to the measurement at 73 ms, which fits the exponential charge growth at points 73.0 through 75.0 ms with a 200 microsec time constant.

**2.C EDGE Detector Data Acquisition**

The EDGE Detector acquired data during delivery in "Snapshot" mode and an example of its response from a delivery can be seen in **Figure 1c.** Snapshot mode stores the recorded samples on the internal buffer instead of sending it to the computer in real time. So it waits for total acquisition of 2048 samples prior to sending them to the computer. This was employed because USB connection currently was not fast enough to transfer the data. Snapshot mode allows capturing data with frequency up to 10 kHz for a limited amount of time. At the highest sampling frequency of 10 kHz, the electrometer can capture 204.8 ms worth of data. The electrometer communication protocol does not have enough bandwidth to send this data in the real time and therefore it is limited to snapshot, which can store M*204.8 ms of data using decimation described earlier. Our measurements set M=4, which results in 400 microsecond spacing between data points and a 819.2 ms of data in the snapshot. The step-like patterns indicate delivery of a radiation pulses. The raw data from the electrometer was processed in MATLAB (Mathworks, Natick, MA) to derive information about each pulse in a delivery and the total accumulated signal. The EDGE response (cumulative and individual pulse) was then compared to the other dosimeters including accumulated dose for film, W1, and IC as well as individual pulse W1 response.

**3. Results**

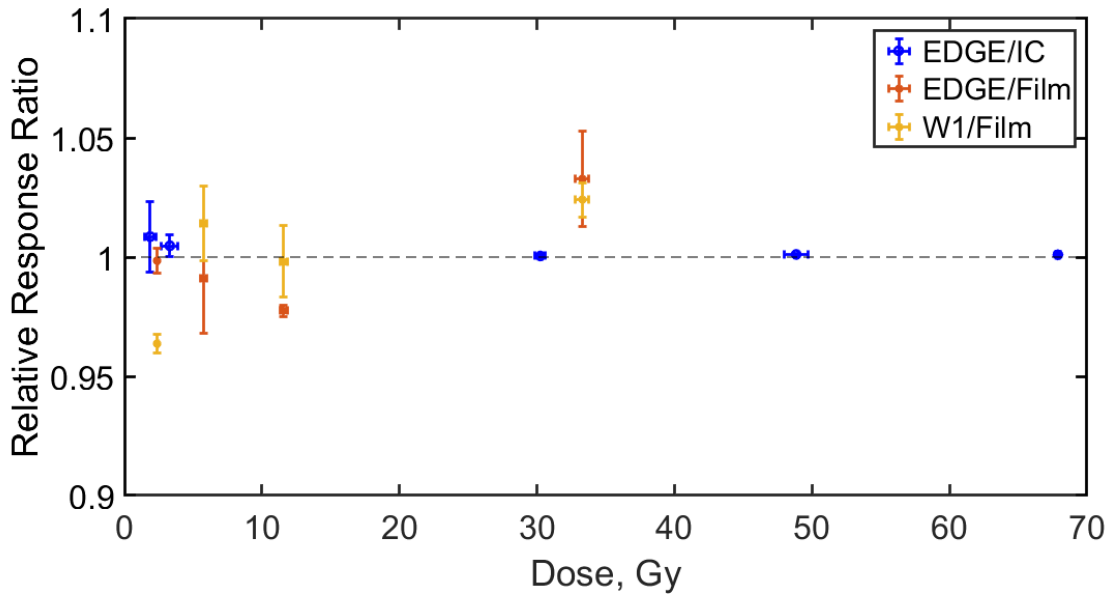

**Figure 2.** W1 and EDGE Detector response in comparison to Gafchromic film and IC with varying dose. The intended dose delivery was controlled based on W1 detector response. Dose was calculated from integrated total dose delivery / number of pulses.

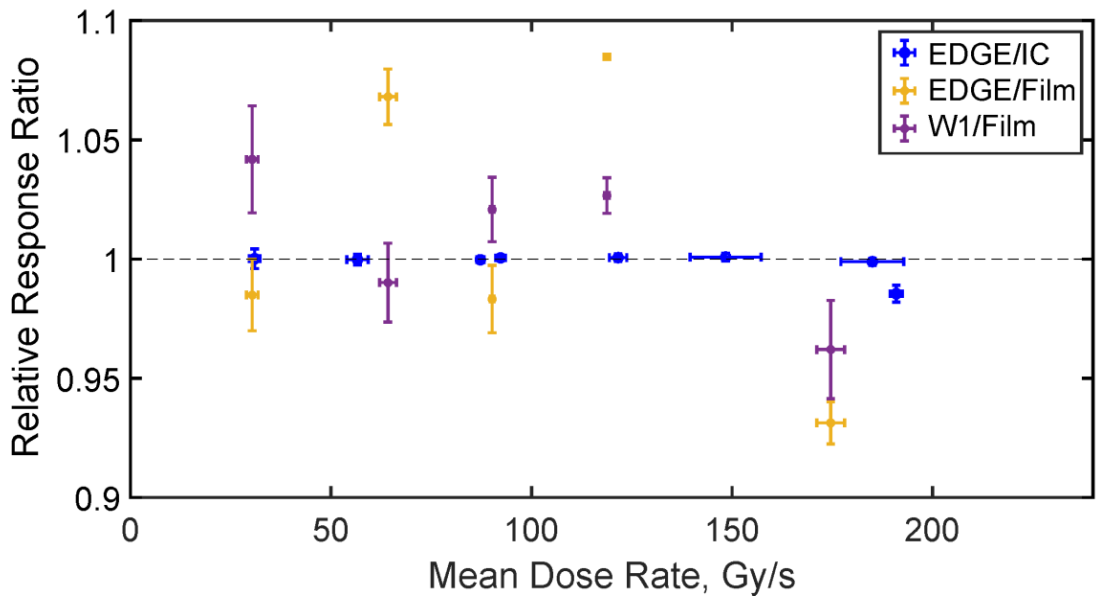

**Figure 3.** W1 and EDGE Detector response in comparison to Gafchromic film and EDGE detector response in comparison to IC for varying mean dose rate. The mean dose rate was varied by changes in the repetition rate of the pulses delivered by the LINAC. The mean dose rate values were calculated from integrated total dose delivery / number of pulses.

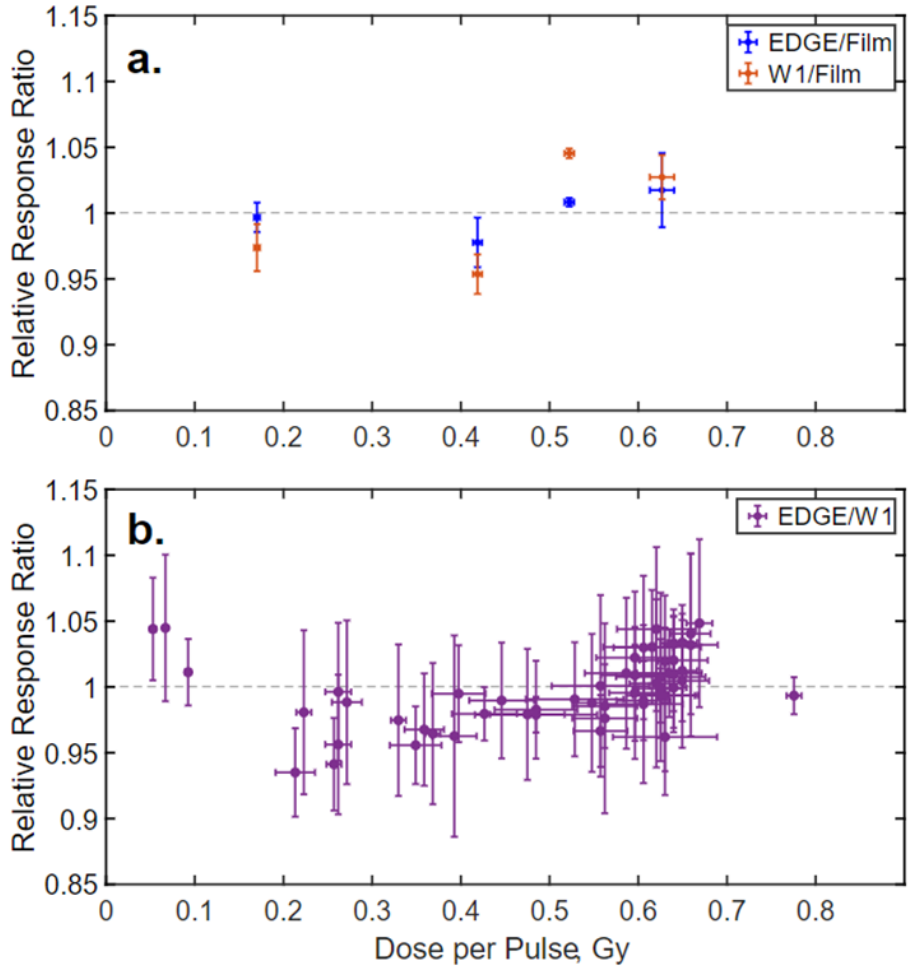

**Figure 4. a.** W1 and EDGE Detector dose per pulse response comparison to Gafchromic film with varying SSD by adjusting the couch's vertical position. The dose per pulse was varied by changing the source to surface distance and adjusting the couch's vertical position. Dose per pulse values were calculated from integrated total dose delivery / number of pulses. **b.** EDGE detector comparison to W1 for individual pulses from 3 deliveries at each of 100 cm SSD position. The varying in individual pulses can be attributed to the LINAC's ramp up period.

**Figure 1c** demonstrates the EDGE detector can measure individual beam pulses to at least 360Hz repetition rate. Following the pulse, the pulse measurement is determined by the difference in values prior to pulse occurrence and the 5$^{th}$ value after the pulse, 8 to 10 time constants (1.6 to 2.0 ms) following the pulse. In comparison, Ashraf[24] et al demonstrated that the W1 detection time was in the order of sub microseconds as the voltage increased for the channels during the beam pulse itself which occurs within ~4 microseconds.

From the data in **Figures 2-4** the authors conclude the independence of EDGE Detector's response and W1 with respect to dose, mean dose rate, and dose per pulse. The error bars for these figures were based on 3 measurements and delivery for each data point. The W1 agreed with film measured dose with varying dose (2.4-33 Gy), dose rate (30 - 180 Gy/s), and dose per pulse (0.17-0.63 Gy/pulse) on average to within -4 to +2%, -4 to +4%, and -5 to +5%, respectively. The EDGE Detector agreed with film measured dose with varying dose, dose rate, and dose per pulse on average to within -3 to +3%, -7 to +8%, and -2 to +2%, respectively. The dose per pulse agreed between the EDGE and W1 detectors on average to within -7 to +5% (0.05-0.78 Gy/Pulse). The EDGE Detector agreed with IC measured dose (1.9 Gy – 70 Gy), and mean dose rate (31 -190 Gy/s) to within +0.8% to 0% and 0% to -1.5% respectively. The IC's $P_{ion}$ and $P_{pol}$ were measured to be 1.04+/-0.02 and 0.99+/-0.01, which is ~1 when considering the uncertainty (associated with the film which provided reference dose in comparison to charge collected). Thus, no correction factor was applied and assumed to be unity.

**Figure 5** and **6** demonstrated the utility of the EDGE Detector for characterizing the electron FLASH beam. The EDGE Detector can measure the PDD of the electron beam (Fig 5), which agreed with the film measured PDD to within 3% (Film PDD was based on 3 measurements per depth). The EDGE also provides per pulse dose or beam output as indicated by the agreement between W1 and EDGE Detector (Fig 6) to within 2% on average**,** including during the variability in the first 30 pulses from the inconsistency in the LINAC pulse output[24,26]. The disagreement in the first three pulses can be attributed to the threshold setting in our electronics that was above the W1 output. Nonetheless the error bars for the response ratio of individual pulses in figure 6b were on average ±4.5% indicating a larger uncertainty in per pulse output measure from EDGE and/or W1.

The EDGE Detector also shows long term stability from dose delivery of at least 4.5 kGy. As shown in **Figure 7**, there has been approximately 2% reduction in response from over 500 deliveries of ~9 Gy.

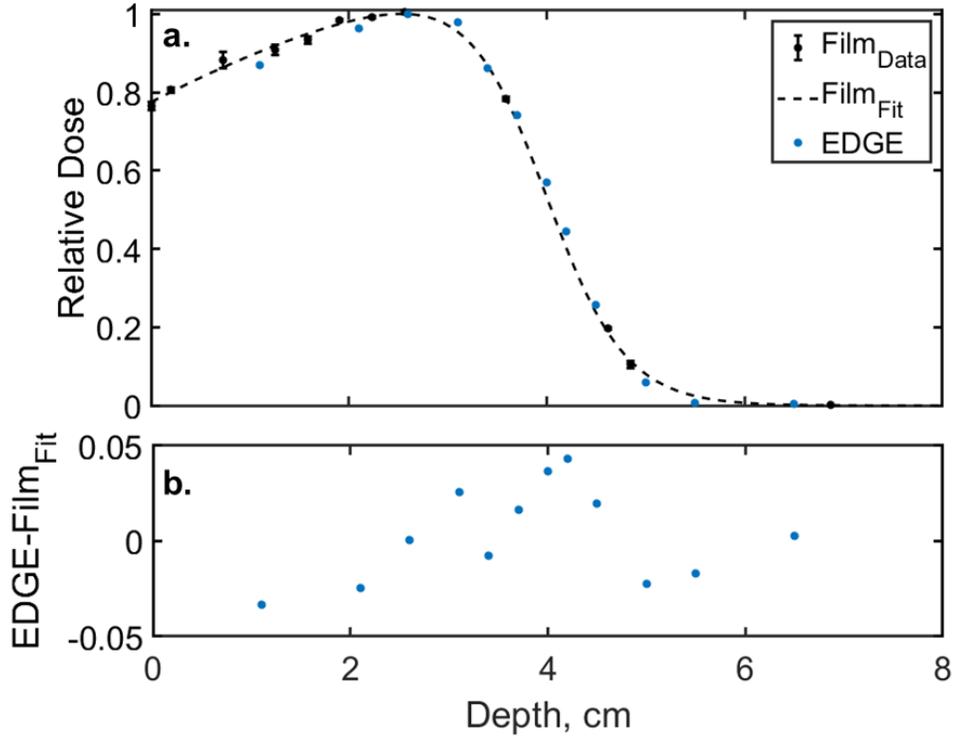

**Figure 5. a.** Percent depth dose curve (PDD) measured by film and EDGE Detector for the 10 MeV UHDR electron beam. Normalized EDGE and film dose values were calculated from integrated total dose delivery / number of pulses. EDGE detector measurement depths ranged from 1.15 to 6.55 cm. **b.** Difference between PDD fitted to Film profile and EDGE Detector measurements.

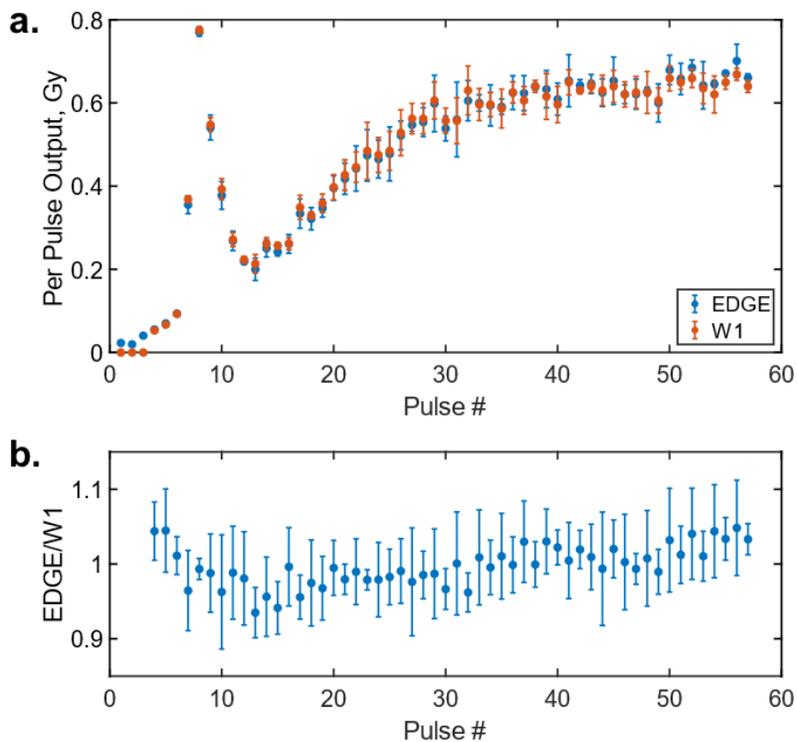

**Figure 6.** a. Relative per pulse beam output from the LINAC delivering UHDR electron beams measured by both the W1 and the EDGE Detector. b. Relative ratio between EDGE detector and W1 scintillator response.

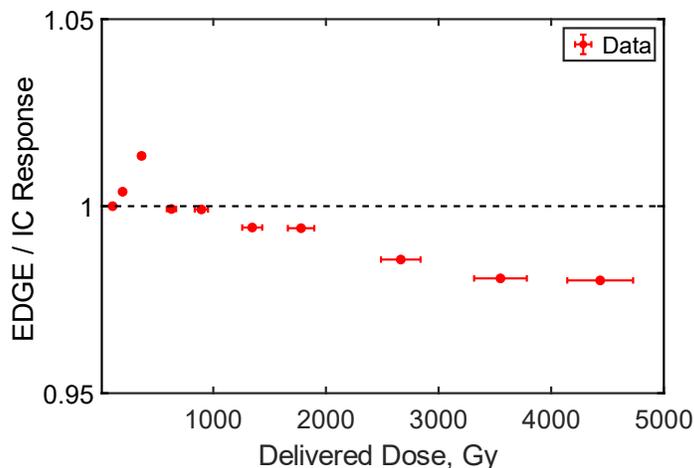

**Figure 7.** Long term sensitivity of the EDGE Detector with respect to the response of the IC.

### 4. Discussion

Characterizing the EDGE Detector with standard dosimeters such as the IC and film and radioluminescence based detector (W1 scintillating fiber) presented some advantages and shortcomings of each. Film has been utilized for characterizing UHDR beam lines, from experimental linac[12] to converted clinical linac[10,26,28] to heavy particle beam lines[4]. This can be attributed to its high spatial resolution and dose rate linearity, requiring a read out via a flatbed scanner post irradiation, thus it accurately provides dose distribution (both lateral and PDD). Nonetheless, figures 2-5 suggests that the uncertainty (from both individual measurements, or uncertainty bars, as well as the range for dose, dose rate, and dose per pulse) can be as much as 7% in comparison to the EDGE Detector and 5% in comparison to the W1 scintillation detector deeming radiochromic film useful as a qualitative dosimeter providing spatial dose distribution but not dose rate or dose per pulse measurements. The uncertainties were in agreement with prior characterization of EBT-XD film[29]. Nonetheless, the agreement between the film and the EDGE Detector for the PDD's demonstrate their energy independence to within 4%.

ICs are considered the gold standard for characterizing and commissioning of clinical beams. One of the advantages over radiochromic film is their precision as it can meet the requirements of annual QA according to AAPM's TG51 report of ensuring beam dose consistency to within 1%. However, ICs experience a significant loss of their collection efficiency in UHDR beams, which is why the reference IC was placed in the bremsstrahlung tail of the beam, where collection efficiency can be considered constant for all measurement setups used in this study. The comparison of the EDGE Detector and the IC, as shown in **Figure 2** and **3**, agreed within 0.4% for both dose and mean dose rate, which means in consequence, that the EDGE detector in conjunction with the modified PCE has a dose and mean dose rate independence within that range. It is worth noting that at lower doses delivered in **Figure 2**, there is a slight deviation away from unity (~1%) and greater uncertainty bar when comparing EDGE and IC, which can be attributed to very low dose within the first few pulses. The mean dose rate was tested with repetition rate (EDGE/IC in **Figure 3**) ranging from 60-360Hz and there was minimal measured dependency. The mean dose rate (repetition rate) independence of the EDGE contrasts the sensitivity drop seen by diode detectors (including EDGE) investigated with conventional integrating electrometer by Jursinic[30] (2013), which may be attributed to the EDGE detector individual pulse measurement with the modified PCE in this work. Another possibility may be due to the range of mean dose rates investigated (0.017-0.1 Gy/s in reference 30 versus ~30-200 Gy/s in this work), which are different by more than 3 orders of magnitude. Ultimately the agreement suggests that the EDGE Detector is precise and can potentially be utilized for annual QA of a FLASH beam line.

Using an IC in the bremsstrahlung tail of the beam as reference can be utilized if the geometry of the set up (e.g., SSD, field size, gantry rotation) is kept constant while other beam output parameters are varied. For example, the LINAC's dose delivery and repetition rate were varied in **Figure 2** and **3**, respectively. However this technique could not be utilized for varying dose per pulse, because dose distribution would change as the source to surface distance is changed[17,18]. So, the ratio of the dose measured at superficial depths (from the EDGE) to the dose at the Bremsstrahlung region would not be consistent. If the dose per pulse is varied by utilizing the pulse structure of the beamline (i.e., pulse width, and pulse height), the IC can be used as the reference for characterizing another dosimeter, like the EDGE Detector.

The W1 scintillation detector benefits from its high temporal resolution (~microsecond), which was utilized to verify the EDGE detector's accuracy in measuring the electron beams dose per pulse. As shown in **Figure 4b**, the two dosimeters agreed within -7 to +5% for individual pulses. The large uncertainty bars in the comparison can be attributed to the variability in measured outputs down to individual pulses from both detectors and variability in the LINAC's output, especially as the output per pulse was reaching stability. It is worth noting, the two dosimeters agreed even when the LINAC output fell to about ~30% of its maximum output during pulses 11-16 as shown in **Figure 6**. Further characterization and optimization of each detector (W1, and EDGE) at UHDR and sub millisecond sampling rate is required for future optimization of LINAC UHDR delivery control and characterization of pulse width analysis using both W1 and EDGE. Pulse width is important to verify UHDR conditions within the pulse. This per pulse dose agreement between 0.05 and 0.8 Gy/pulse (range of dose per pulse in **Figure 4b**) potentially implies dose per pulse independence of the EDGE and W1 detectors. However, the W1 exhibits more radiation damage as there was 16% reduction in response for the first kGy of dose and then response drops by 7% per kGy measured by Ashraf et al[24], while in this study the EDGE Detector exhibited a reduction in response ~0.4% per kGy as shown in Figure 7. Nonetheless at conventional dose rates the sensitivity reduction of W1 was at most 0.28% per kGy[21]. W1 degradation from single deliveries of ~60 pulses (~27 Gy) were ~0.4% in this study. Alternatively to further confirm the dose per pulse independence of the diode and per-pulse temporal characteristics of the electron beam, the diodes response can be compared to a modified sensitivity diamond detector such as one from Kranzer et al.[20] which achieved 6.5 Gy/pulse and could potentially resolve pulses up to 500 Hz, with insignificant sensitivity change over 1MGy.

The EDGE Detector seems to exhibit the advantages of the other dosimeters used to characterize it in this study. Its detector in previous studies[31] characterized small fields indicating high spatial resolution like film. In this study the detector provided the PDD curve and its agreement with film implies energy independence as there is energy loss of electrons with greater depths, further confirmed with independence exhibited for MV x-ray beams[32]. The detector may in the future be utilized for lateral profiles of UHDR small fields and patient specific cutouts. Nonetheless the detector will require characterization under small fields, directional dependence, and temperature dependency under UDHR conditions. While IC is often utilized for a machine's annual QA due to its high precision and reproducibility[33], the EDGE Detector also agreed with the IC to within <0.4% difference. It also benefits from fast readout like that of an IC, without the concern of correction for the ion recombination at UHDR. It may also be utilized as a dosimetry tool *in vivo* during FLASH treatment of small or large animals like that used by Ashraf et al[24] with the W1 scintillation detector. The 0.1 ms sampling rate can ensure resolving individual pulses of the UHDR electron beam similar to the W1 scintillator, with reduced damage from radiation. Thus, the EDGE Detector may be a candidate for controlling and stopping UHDR beam delivery based on prescribed dose. However, the EDGE detector in its current state cannot provide intra-pulse dose rate as there is approximately a ~1ms time for charge to accumulate, and the individual pulses have ~4 microsecond width. The charge accumulation time, shown in the zoomed in plot of Figure 1c, can be attributed to the buffer box in the circuit for measuring the charge which is composed of an RC circuit (2 KΩ resistance and 100nF capacitor) that results in a time constant of 200 microseconds. If EDGE detectors can be placed laterally in four cardinal directions closer to the beam source (e.g., before the applicator or cutout), ideally using an electrometer with 4 channels that has sensitivity for UHDR beams, the beam may potentially be stopped based on changes in symmetry and shift analogous to the monitor chambers controlling the beam in conventional clinical LINACs. Nonetheless, this study focused on one N-type diode dosimeter, and it motivates future studies exploring effectiveness of other diode detectors (N-type, P-type, PIN type) under UHDR conditions.

## 5. Conclusion

This study demonstrated that the EDGE diode detector can be used for characterization of UHDR electron beams. It is linear with dose, dose per pulse, and mean dose rate with an average decline in sensitivity of 0.4% per kGy. It exhibited several of the benefits of other dosimeters including fast read-out, and precision comparable to the gold standard ionization chambers. It recorded real time individual pulse beam delivery like scintillation-based detectors, with high spatial resolution though at a courser resolution than radiochromic film. It exhibited reduced radiation damage compared to the W1 scintillation detector, but still its sensitivity fade must be considered when used over long time periods.

The EDGE Detector's submillisecond temporal resolution and energy independence at the MeV energy range suggests potential to characterize electron beams spatially and at individual pulses. Its utility can potentially be expanded beyond small field dosimetry, to provide *in vivo* dose during FLASH experiments, as well as monitor beam profiles and stop beam down to individual pulses. This will require further development of the technology, implementation into a beam monitoring and stopping device, and verification of minimal perturbation on the beam delivered to the biological subjects.

**6. Acknowledgments:** This work was supported by the Norris Cotton Cancer Center seed funding through core grant P30 CA023108 and through the shared irradiation service, as well as through seed funding from the Thayer School of Engineering, and from grant R01 EB024498 and U01 CA260446. Department of Medicine Scholarship Enhancement in Academic Medicine (SEAM) Awards Program from the Dartmouth Hitchcock Medical Center and Geisel School of Medicine also supported this work.